\documentclass[aps,prl,preprint,groupedaddress]{revtex4}

\usepackage{amsmath}

\usepackage{slashed}

\usepackage{graphicx}

\begin{document}


\title{Discovery Prospects of an Almost Fermiophobic $W^\prime$ in the
Three-Site Higgsless Model at the LHC}


\author{Fabian Bach}
\email[]{fabian.bach@physik.uni-wuerzburg.de}
\author{Thorsten Ohl}
\email[]{ohl@physik.uni-wuerzburg.de}
\affiliation{Institut f\"ur Theoretische Physik und Astrophysik,
Universit\"at W\"urzburg, Hubland Nord, 97074 W\"urzburg, Germany}


\date{\today}

\begin{abstract}
  In extensions of the Standard Model with compactified extra
  dimensions, perturbative unitarity of the longitudinal gauge bosons
  is maintained through the contribution of heavy KK excitations of
  the gauge fields, without the necessity of introducing a Higgs
  field. The \emph{Three-Site Higgsless Model} represents a minimal
  approach in this respect, containing just one extra set of heavy
  gauge bosons $Z^\prime$/$W^{\prime\pm}$ in the spectrum. While the
  $Z^\prime$ can have robust couplings to SM fermions and hence may be
  detected within the first \mbox{$1$--$20\,$fb$^{-1}$} of LHC data
  (\mbox{$\sqrt{s}=14\,$TeV}), the coupling of the $W^\prime$ to light
  fermions is suppressed and depends on the model
  parameters. Expanding on previous parton level studies, we
  determine discovery thresholds of the $W^\prime$ in $s$-channel
  Drell-Yan~production at the LHC for masses $m_{W^\prime}=380$, $500$
  and \mbox{$600\,$GeV}, combining analyses of the semileptonic final
  states $\ell\ell jj$, $\ell\nu jj$ and the leptonic final state
  $\ell\nu\ell\ell$ (\mbox{$\ell=e,\mu$}) including fast detector simulation.
\end{abstract}


\maketitle


\section{Introduction}\label{intro}

The \emph{Standard Model of Particle Physics} (SM) has been an
extremely successful theory over the past decades, consistently
explaining and predicting many different experimental results in high
energy physics with unprecedented precision.  While the
experimental evidence for a breaking of the underlying local
symmetry is unambiguous, the dynamics of this spontaneous
breaking of the electroweak symmetry (EWSB) has not been
accessible experimentally so far and remains an open question
until now. Indeed, the \emph{Large Hadron Collider}~(LHC)
at CERN has been designed to reach the
electroweak breaking scale in hard parton scatterings and
illuminate this yet unresolved question experimentally.

On the theoretical side, numerous approaches exist in the literature
to implement a dynamical mechanism of symmetry breaking in the SM
which also keeps the scattering of longitudinal gauge bosons unitary
at high energies
above~$1\,$TeV~\cite{Dicus:1992vj,Cornwall:1973tb,Cornwall:1974km,Lee:1977yc,Lee:1977eg,Chanowitz:1985hj},
among which the introduction of a fundamental scalar Higgs
field~\cite{Higgs:1964ia,Higgs:1964pj,Lee:1977eg,Lee:1977yc} is only
the most simple and straightforward one, albeit with some
much-discussed theoretical drawbacks such as the quadratic
dependence on the ultraviolet completion
known as the \emph{hierarchy problem}. Other approaches
are, for example, technicolor models with an additional strongly
interacting sector~\cite{Weinberg:1975gm,Dimopoulos:1979es}, whose
meson-like bound states play the role of the symmetry-breaking
scalars, or models with one or more additional space-time dimensions
compactified on the electroweak length
scale~\cite{Randall:1999ee,Maldacena:1997re,Maldacena:1999fi}, where
symmetry breaking can be accounted for by non-trivial ground state
configurations of the additional gauge field components.

Indeed, in the case of five dimensions with the geometry of a 5D
\mbox{Anti-de~Sitter} space, these different types of models turn out
to be related to each other by a duality commonly addressed as AdS/CFT
correspondence~\cite{Maldacena:1997re,Maldacena:1999fi}. The
Three-Site Higgsless Model (3SHLM)~\cite{Chivukula:2006cg} can be
counted among this class of theories as a maximally
deconstructed~\cite{ArkaniHamed:2001ca,Hill:2000mu} limit, thus
representing an effective low energy theory parametrizing only the
leading contributions of a more general higgsless 5D theory as a UV
completion~\cite{Georgi:1989gp,Georgi:1989xy,Bando:1987br,Harada:2003jx}. More
explicitly, maximal deconstruction means that the extra dimension is
discretized on just three sites, giving rise to the inclusion of only
the first Kaluza-Klein (KK) excitations of the SM fields along the new
dimension into the particle spectrum. While electroweak precision
tests (EWPT) severely constrain the mass scale of the heavy fermions
to \mbox{$\gtrsim 2\,$TeV}, the masses of the heavy gauge bosons
$Z^\prime$/$W^{\prime\pm}$ can be consistently chosen as low as
\mbox{$380\,$GeV}~\cite{Chivukula:2006cg,Abe:2008hb,Abe:2009ni},
whereas an upper limit around \mbox{$600\,$GeV} is dictated by the
necessity to unitarize the scattering amplitudes among longitudinal
gauge bosons. Although ideal delocalization allows to implement an exact vanishing
of the $W^\prime$ coupling to SM fermions $g_{W^\prime ff}$ at the
tree level~\cite{Chivukula:2006cg}, the results of a one-loop
analysis~\cite{Abe:2008hb} actually favour small but finite couplings.

In view of the major LHC experiments currently gaining precision and
improving bounds on new physics signatures literally every month, there
is an ongoing interest in higgsless models predicting new gauge bosons
with strongly suppressed couplings to SM fermions, making them hard to
access experimentally.
Even though there already exist quite a few studies dealing with the
discovery prospects of such heavy gauge bosons
~\cite{Ohl:2008ri,He:2007ge,Bian:2009kf,Han:2009em,Han:2009qr,Asano:2010ii,
Brooijmans:2010tn,Speckner:2010zi,Accomando:2011xi}
(the most recent one~\cite{Accomando:2011xi} discussing the LHC discovery
reach of heavy charged gauge bosons $W^\pm_{\text{1,2}}$ in the
Four-Site Higgsless Model, whose couplings are less constrained than in the 3SHLM),
most of them are restricted to the parton level, which does not
account for---possibly large---effects due to the detector
response. Discovery limits for the $Z^\prime$ including detector
effects have recently been reported elsewhere~\cite{Abe:2011qe}, but
the discovery of the $W^\prime$ is crucial to distinguish the 3SHLM in
the one-loop scenario from a generic $Z^\prime$ with suppressed
couplings.  This paper is focussing on the sensitivity to the
$W^\prime$ in $s$-channel production. We compare the most promising
final states $\ell\ell jj$, $\ell\nu jj$ and $\ell\nu\ell\ell$
(\mbox{$\ell=e,\mu$}) of a $W^\prime$ decaying predominantly into
intermediate pairs of SM gauge bosons $WZ$ in order to extract the
most sensitive one at the detector level, also including an assessment
of the method studied in~\cite{Ohl:2008ri} at parton level to
distinguish the nearly degenerate $W^\prime$ and $Z^\prime$ resonances
in the $\ell\nu jj$ channel. Note that, apart from the specific case
discussed here, this method is in principle applicable to a more general
class of signal patterns where resonances lie close together, especially
when data analysis is further complicated, e.\,g.~by information loss due
to invisible particles in the final state.

This article is organized as follows: Section~\ref{tshlm} is devoted
to a brief review of the construction principles and parameter space
of the 3SHLM, while in section~\ref{wsearch} we point out the
generation and detector simulation of the analyzed samples and present
sensitivities and discovery prospects in the different channels
mentioned. Finally, a summary and detailed discussion of the results
can be found in section~\ref{summ}.

\section{The Three-site Higgsless Model}\label{tshlm}

The 3SHLM~\cite{Chivukula:2006cg}
can be understood as an effective low-energy approximation
of a 5D theory with one extra space dimension $y$ of the size of the
EWSB scale and a gauged 5D $\mathbf{SU}(2)$ bulk symmetry broken to
a $\mathbf{U}(1)$ on one of the branes.  In addition, $y$ is
maximally deconstructed, i.\,e.~discretized to three sites only,
so that only the ground states and the first KK excitations of
the matter and gauge fields along $y$ remain in the spectrum.
In the effective 4D theory, this deconstruction corresponds to an extended
electroweak gauge group
\begin{equation}\label{eq:ggroup}
 \mathbf{SU}\left(2\right)_0 \times \mathbf{SU}\left(2\right)_1 \times \mathbf{U}\left(1\right)_2,
\end{equation}
where the chiral $\mathbf{SU}(2)_L \times \mathbf{U}(1)_Y$
symmetry of the SM is realized via boundary conditions breaking the full symmetry
on the brane sites 0 and 2.
In the limit $g_{0/2}\ll g_1$, the gauge couplings can be approximately
identified as the electroweak SM couplings $g_0\sim g$ resp. $g_2\sim g^\prime$,
and any new physics contribution from \mbox{site~1} will be suppressed as
$g_{0/2}/g_1$.

In this setup, EWSB can be implemented by a non-trivial vacuum configuration
of the $y$-component of the 5D gauge field $A_M(x)$ with \mbox{$M=0,1,2,3,y$}.
In the deconstructed 4D picture, the component $A_y(x)$, i.\,e.~the gauge
connection between the sites, is mediated by two independent
$\mathbf{SU}(2)$-valued Wilson line fields $\Sigma_{0/1}(x)$ which transform
bi-unitarily under gauge transformations as
\begin{equation}
 \Sigma_0 \to U_0 \Sigma_0 U_1^\dagger\;,\qquad \Sigma_1 \to U_1 \Sigma_1 e^{-i\theta\sigma_3/2}
\end{equation}
with gauge group elements $U_i\in\mathbf{SU}(2)_i$ and $e^{-i\theta\sigma_3/2}\in\mathbf{U}(1)_2$.
Choosing the minimal nonlinear sigma representation for the Wilson lines,
the $\Sigma_i(x)$ become unitary and EWSB is realized by the vacuum
expectation values
\begin{equation}\label{eq:vev}
 \langle \Sigma_{0/1} \rangle=\sqrt{2}v
\end{equation}
(this symmetry breaking pattern is very similar to the BESS
model~\cite{Casalbuoni:1985kq}).

Together with eqn.~(\ref{eq:vev}), the kinetic terms of the $\Sigma_i(x)$
generate mass terms for the gauge fields, which upon diagonalization, and
in the limit $g_{0,2}\ll g_1$ mentioned above, lead
to a massless photon $A$ and a light set of mass eigenstates $W^\pm$/$Z$
mostly localized on the boundary branes (to be identified with the SM
gauge bosons) as well as a heavy, almost degenerate set $W^{\prime\pm}$/$Z^\prime$
of mass $m^\prime\sim g_1v$ localized on the bulk \mbox{site~1}.
The allowed range of $m^\prime$ is
\begin{equation}
 380\,\text{GeV} \lesssim m^\prime \lesssim 600\,\text{GeV}\;,
\end{equation}
where the lower bound comes from the LEP2 measurement of the triple
gauge boson coupling $g_{ZWW}$~\cite{Hagiwara:1986vm,:2005ema} and the
upper bound is due to the requirement that the heavy gauge bosons be
light enough to delay the violation of unitarity in longitudinal gauge
boson scattering.

5D bulk fermion fields are broken up by deconstruction
into independent fermion fields $\Psi_i$ at each \mbox{site~$i$}, whereas
the chirality of the SM is enforced by boundary conditions, i.\,e. by
requiring $\Psi_0$ to be a left-handed doublet under $\mathbf{SU}(2)_0$ and
the $\Psi_2^{u/d}$ to be right-handed and singlets under all the $\mathbf{SU}(2)_i$, while the
bulk fermions $\Psi_1$ remain fully left-right symmetric doublet representations
of the $\mathbf{SU}(2)_1$. In addition, the $\Psi_{0/1}$ carry $\mathbf{U}(1)_2$
charges equal to the hypercharges of left-handed SM fermions, whereas the
$\mathbf{U}(1)_2$ charges of the $\Psi_2^{u/d}$ are equal to the right-handed
SM hypercharges. Mass terms are then generated by a combination of Yukawa
couplings to the $\Sigma$ fields and a gauge invariant Dirac mass term
$M\overline{\Psi}_{1L}\Psi_{1R}$ which normalizes the overall mass scale:
\begin{widetext}
\begin{equation}\label{eq:yuk}
 \mathcal{L}_{\text{mass}} =  M \left[ \epsilon_L \overline{\Psi}_{0L} \Psi_{1R} + \overline{\Psi}_{1L} \Psi_{1R} + \overline{\Psi}_{1L} \left(\begin{array}{cc} \epsilon^u_{R}\\ & \epsilon^d_{R} \end{array} \right) \left(\begin{array}{c} \Psi_{2R}^{u}\\ \Psi_{2R}^{d}\end{array}\right) \right] + \text{h.c.}\; ,
\end{equation}
\end{widetext}
where the vev's of the $\Sigma_i$, \mbox{eqn.~(\ref{eq:vev})}, have been
inserted and a sum over the three flavor generations is implied. In this
parametrization, \mbox{$M\gtrsim 2\,$TeV} is a universal heavy fermion mass scale,
with the lower bound coming from heavy fermion loops contributing to the $W$
propagator and hence affecting EWPT~\cite{Chivukula:2006cg}, and the
$\epsilon_i\sim v/M$ are dimensionless delocalization parameters
mixing fermions from adjacent sites.

Again, after diagonalization one finds a heavy set of fermions mostly
living on the bulk site and a light set localized on the boundaries which
is identified with the SM fermions. The SM flavor structure is incorporated in
a minimal way, i.\,e.~by fixing the matrices $\epsilon^{u/d}_{R}$
according to flavor phenomenology while keeping $M$ and $\epsilon_L$ as universal,
free parameters of the model (together with $m^\prime$). However, $\epsilon_L$ is strongly correlated with $m^\prime$
in order to keep the model consistent with EWPT. 
More explicitly, the coupling $g_{W^\prime ff}$ of the $W^\prime$ to SM
fermions is straightforwardly given by the gauge couplings times the overlap
of the wave functions of mass states on the respective sites
\begin{equation}\label{ideloc}
 g_{W^\prime ff} = g_0 \left( f^0_L \right)^2 v^0_{W^\prime} + g_1 \left( f^1_L \right)^2 v^1_{W^\prime}\;,
\end{equation}
where the $f^i_L$ and $v^i_{W^\prime}$ are found in the diagonalization
of the mass matrices and relate the mass states of the light left-handed fermions
and the $W^\prime$, respectively, to the corresponding interaction eigenstates
on \mbox{site $i$}, with $f^0_L/f^1_L\sim\epsilon_L$ and
$v^0_{W^\prime}/v^1_{W^\prime}\sim m_W/m^\prime$ to leading order.
At tree level, \mbox{eqn.~(\ref{ideloc})} is required to give exactly zero
in order to protect the electroweak precision parameters from $W^\prime$
contributions, thus imposing a strict relation between $\epsilon_L$ and
$m^\prime$---\mbox{a setup} referred to as
\emph{ideal delocalization}~\cite{Chivukula:2005xm}
in~\cite{Chivukula:2006cg}.
However, as pointed out in~\cite{Abe:2008hb},
at one-loop level ideal delocalization turns out to be ruled out at
\mbox{$95\,\%$~c.l.} in favor of a small but non-vanishing coupling
$g_{W^\prime ff}$ of the $W^\prime$ to the SM fermions, so that the search
for a $W^\prime$ resonance in the $s$ channel at the LHC will be sensitive
not only to $m^\prime$ but also to $\epsilon_L$ and to $M$. In the present
study, we will essentially rely on the \mbox{$95\,\%$~c.l.} allowed regions
in the $M$--$\epsilon_L$ parameter plane given in~\cite{Abe:2008hb} for
fixed heavy gauge boson masses $m^\prime=380$, $500$ and \mbox{$600\,$GeV}.

\section{Discovery Potential of the $W^\prime$ at the LHC}\label{wsearch}

In this section, we present our results of a Monte~Carlo-based feasibility
study of the search for the resonance of a 3SHLM-like $W^\prime$
boson which is produced in the non-ideal delocalization setup via $s$ channel
quark annihilation. It predominantly decays into a pair of SM gauge bosons
\mbox{$W^\prime\to WZ$}, subsequently decaying into four-fermion final states,
of which $\ell\ell jj$, $\ell\nu jj$ and
$\ell\nu\ell\ell$ ($\ell=e,\mu$) are most promising due to moderate
($\ell\ell jj$, $\ell\nu jj$) or absent ($\ell\nu\ell\ell$) QCD backgrounds,
whereas the purely hadronic final state $jjjj$ was not considered here due to
the huge QCD background. The data generation and simulation setup of the
study is designed to reproduce LHC-like proton--proton collisions at
\mbox{$\sqrt{s}=14\,$TeV} and to approximately imitate the
ATLAS~\cite{Aad:2009wy} detector response via a fast detector simulation
with smearing effects.

\subsection{Data Generation and Detector Simulation}

All signal and background samples have been produced with version~2 of
the parton-level
Monte~Carlo event generator WHIZARD~\cite{Kilian:2007gr}. All final states
containing up to four particles were generated using full tree level matrix
elements including all off-resonant diagrams and irreducible backgrounds.
This already covers most of the dominant QCD backgrounds coming from $W$/$Z$
plus one or two colored partons in the semileptonic channels. For the
additional \mbox{$t\overline{t}$~background} in the $\ell\nu jj$ channel, the
computational complexity has been reduced by selecting only those diagrams which
contain two $b$ quarks and two $W$ boson propagators decaying semileptonically.

All WHIZARD output was then processed with PYTHIA~\cite{Sjostrand:2006za}
for showering and hadronization. Finally, detector effects have been accounted
for using the fast detector simulation DELPHES~\cite{Ovyn:2009tx} with the
default ATLAS chart shipped with the package. Calorimeter jets have been
reconstructed with the anti-$k_T$ algorithm~\cite{Cacciari:2008gp} and cone
size \mbox{$R=0.4$}. Also $b$-tagging has been applied with an assumed
efficiency of $0.6$~\cite{Aad:2009wy} in order to suppress the
\mbox{$t\overline{t}$~background} in the $\ell\nu jj$ channel.

\subsection{The $\ell\ell jj$ Channel}\label{lljj}

The $\ell\ell jj$ channel is the most straightforward one, because all final
state objects are visible in the detector, so that their momenta can in
principle be uniquely determined from detector data. Moreover, this channel
contains no $Z^\prime$ signal, thus conveying a clean signature to
discriminate the one-loop \mbox{scenario~\cite{Abe:2008hb}} from the ideal
delocalization \mbox{scenario~\cite{Chivukula:2006cg}}, where the $W^\prime$
resonance should vanish completely.

\begin{figure*}
 \includegraphics[scale=0.35]{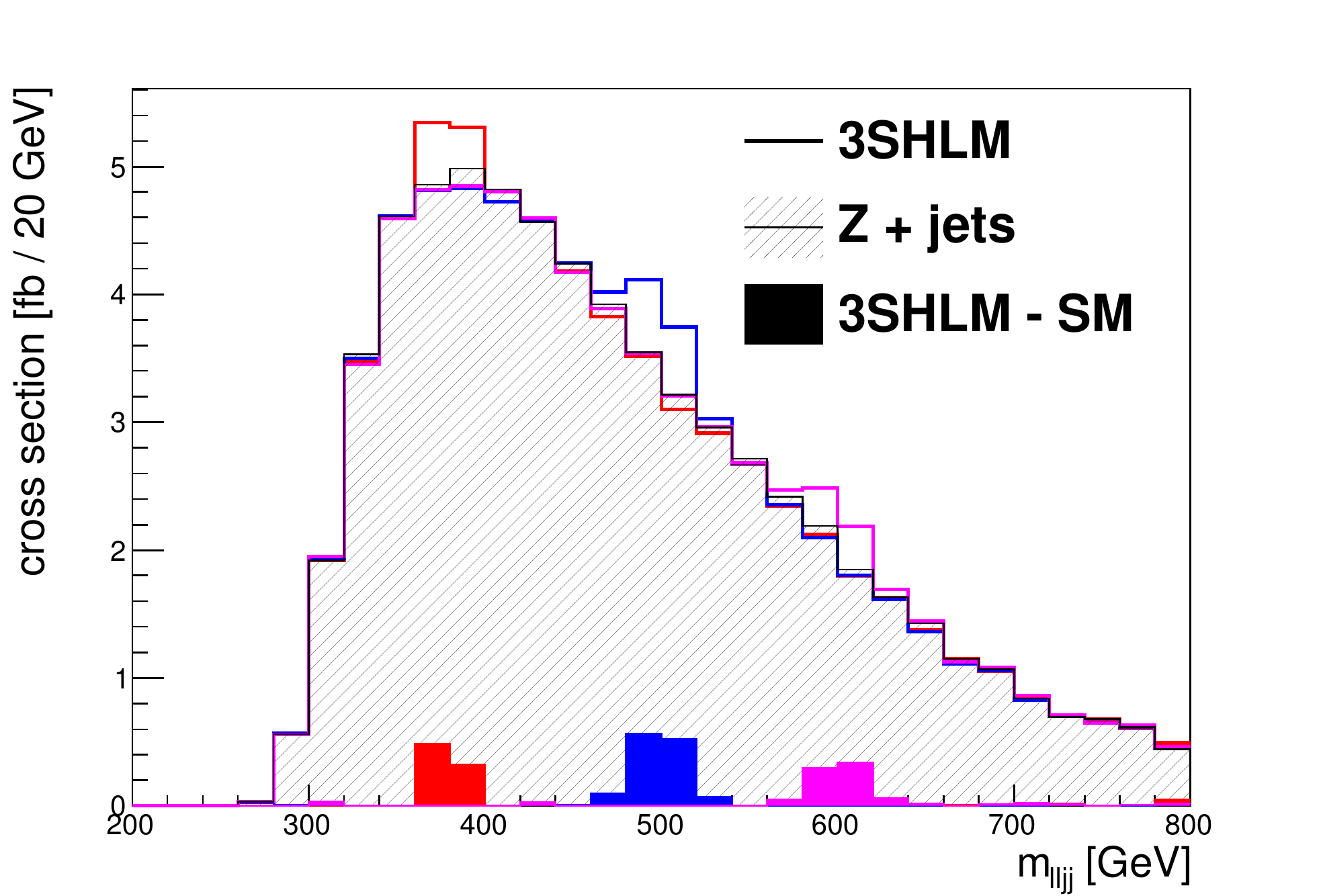}
 \includegraphics[scale=0.35]{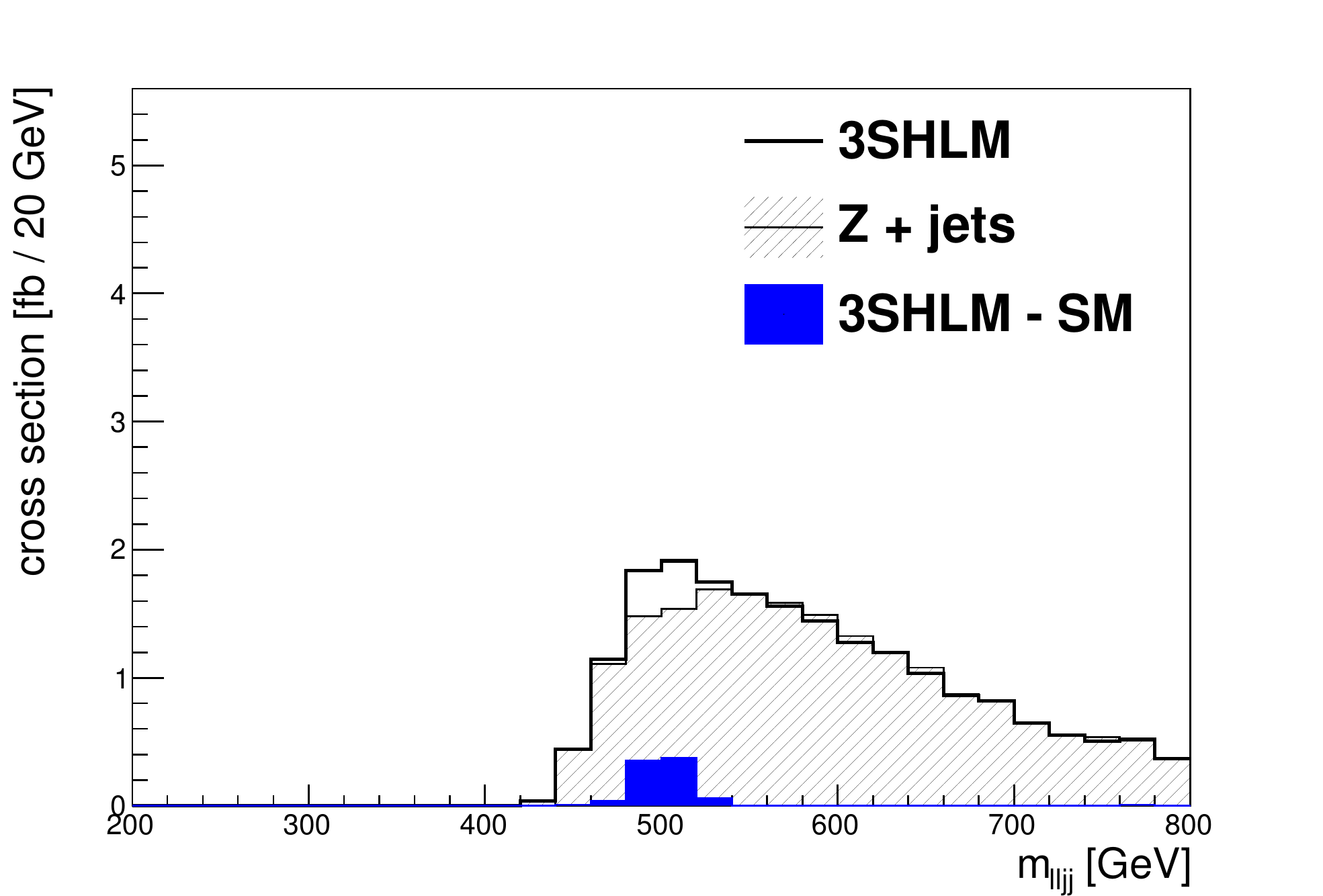}
 \caption{Total invariant mass distribution of the final state $\ell\ell jj$
after all cuts without (l.h.s) and including (r.h.s.) the $p_T$ cut on the
SM gauge boson momenta, with maximal $g_{W^\prime ff}$ allowed in the
one-loop scenario and \mbox{$m^\prime=380$} (red), 500 (blue) and
\mbox{$600\,$GeV} (pink).\label{lljj_hW}}
\end{figure*}

For event selection, we require exactly two isolated, oppositely
charged leptons as well as at least two hadronic jets all passing the
kinematic cuts
\begin{equation}\label{kincuts}
 p_T > 50\,\text{GeV} \qquad \text{and} \qquad \left|\eta\right| < 2.5
\end{equation}
together with a cut on the invariant di-lepton mass within
$\sim 2\,$FWHM around the $Z$ mass:
\begin{equation}\label{Zcut_ll}
 \left| m_{ee} - m_Z \right| < 5 \,\text{GeV} \quad \text{resp.} \quad \left| m_{\mu\mu} - m_Z \right| < 10\,\text{GeV} \;.
\end{equation}
In order to further reduce QCD background, additional kinematic cuts are
applied to the hadron jets, which restrict the allowed phase space to the
topology associated with the decay of two strongly boosted SM gauge bosons.
The corresponding requirements are to find a jet pair with an invariant
mass within a cut window around the $W$ mass and a relatively small enclosed
angle,
\begin{equation}\label{Wcut_jj}
 \left| m_{jj} - m_W \right| < 10 \,\text{GeV} \quad \text{and} \quad \Delta R\left( j,j \right) < 1.3 \;.
\end{equation}
Finally, with the two lepton momenta and the momenta of the di-jet resonance,
a $p_T$ \mbox{cut~\cite{Abe:2011qe}} as well as a back-to-back cut is applied
on the reconstructed gauge boson momenta associated with the lepton pair and
the jet pair:
\begin{align}\label{WZ_btb}
 p_T^{W/Z} &> \left\{ \begin{array}{ll} 150\,\text{GeV}, & m^\prime=380\,\text{GeV} \\ 200\,\text{GeV}, & m^\prime=500\,\text{GeV} \\ 250\,\text{GeV}, & m^\prime=600\,\text{GeV} \end{array} \right. \;, \nonumber \\
 \Delta R\left( W,Z \right) &> \quad\, 2.0 \;.
\end{align}
The signal observable is then given by the total invariant mass
$m_{\ell\ell jj}$ of all four 4-momenta in the final state, where a veto is
imposed on b-tagged jets passing \mbox{$p_T>25\,$GeV} and
\mbox{$\left|\eta\right|<2.5$} cuts in order to further reduce background.
(Particularly, $t\overline{t}$ background can be neglected when requiring two
hard leptons reconstructing a $Z$ in combination with a veto on $b$ jets and
missing transverse momentum.)

Significances are obtained by adding the reducible background samples to
the signal as well as the SM contrast sample and evaluating the excess
events in the resonance region normalized by the background fluctuation,
\begin{equation}\label{def_sign}
 s = \frac{N_{\text{3SHLM}}-N_{\text{SM}}}{\sqrt{N_{\text{SM}}}}\;,
\end{equation}
where the resonance region is chosen symmetrically around $m^\prime$.
Note that $m^\prime$ is assumed to be known within the 3SHLM from a discovery
of the practically degenerate $Z^\prime$, which must necessarily have taken place
before any sensitivity to the $W^\prime$ in the $s$ channel can be expected
due to the much larger coupling. The width of the resonance window is
adjusted such that the value for the significance minus its statistical
uncertainty is optimized. As illustrated in \mbox{fig.~\ref{lljj_hW}},
however, the $\ell\ell jj$ channel suffers greatly from the tiny signal
together with large QCD backgrounds, giving \mbox{e.\,g.} a $5\sigma$
discovery threshold of \mbox{$\int L\approx150\,$fb$^{-1}$} at the most
optimistic point of the parameter space (\mbox{$m^\prime=500\,$GeV}, max.
$g_{W^\prime ff}$ allowed), so that the possibility of a discovery in this
channel remains questionable on the basis of this study, at least with more
or less realistic assumptions on the total integrated luminosity to be
delivered by the LHC.

\subsection{The $\ell\nu jj$ Channel}\label{lvjj}

This channel is somewhat more involved for a $W^\prime$ search than the
previous one due to the missing kinematic information of the neutrino which
escapes detection. Furthermore, the $W^\prime$ signal is superimposed with
a large degenerate $Z^\prime$ resonance (\mbox{cf.~\cite{Abe:2011qe}}), so
that the possibility to extract the $W^\prime$ is closely linked to the
ability to distinguish the heavy SM gauge boson resonances in the
di-jet system~\cite{Ohl:2008ri}. These issues are addressed below in
due order.

The selection criteria for this channel are very similar to those described
in \mbox{sec.~\ref{lljj}} for the $\ell\ell jj$ channel: we basically require
at least two hard jets passing the kinematic \mbox{cuts (\ref{kincuts})}
together with an invariant mass cut
\begin{equation}
 m_W - 20\,\text{GeV} < m_{jj} < m_Z + 20\,\text{GeV}\;,
\end{equation}
which accounts for the fact that in the di-jet mass a small $Z$ resonance
coming from the $W^\prime$ is added to a large $W$ resonance coming from a
$Z^\prime$. (The cut window is increased compared \mbox{to~(\ref{Wcut_jj})}
because of the discrimination procedure described below.) In addition, we
require exactly one hard lepton passing the kinematic
\mbox{cuts (\ref{kincuts})} as well as missing transverse energy
$\slashed{E}_T$ larger than the $p_T$ cut \mbox{in (\ref{kincuts})},
together with a $b$-jet veto as described in \mbox{section~\ref{lljj}}
in order to suppress the large $t\overline{t}$ background in this channel.

\begin{figure*}
 \includegraphics[scale=0.35]{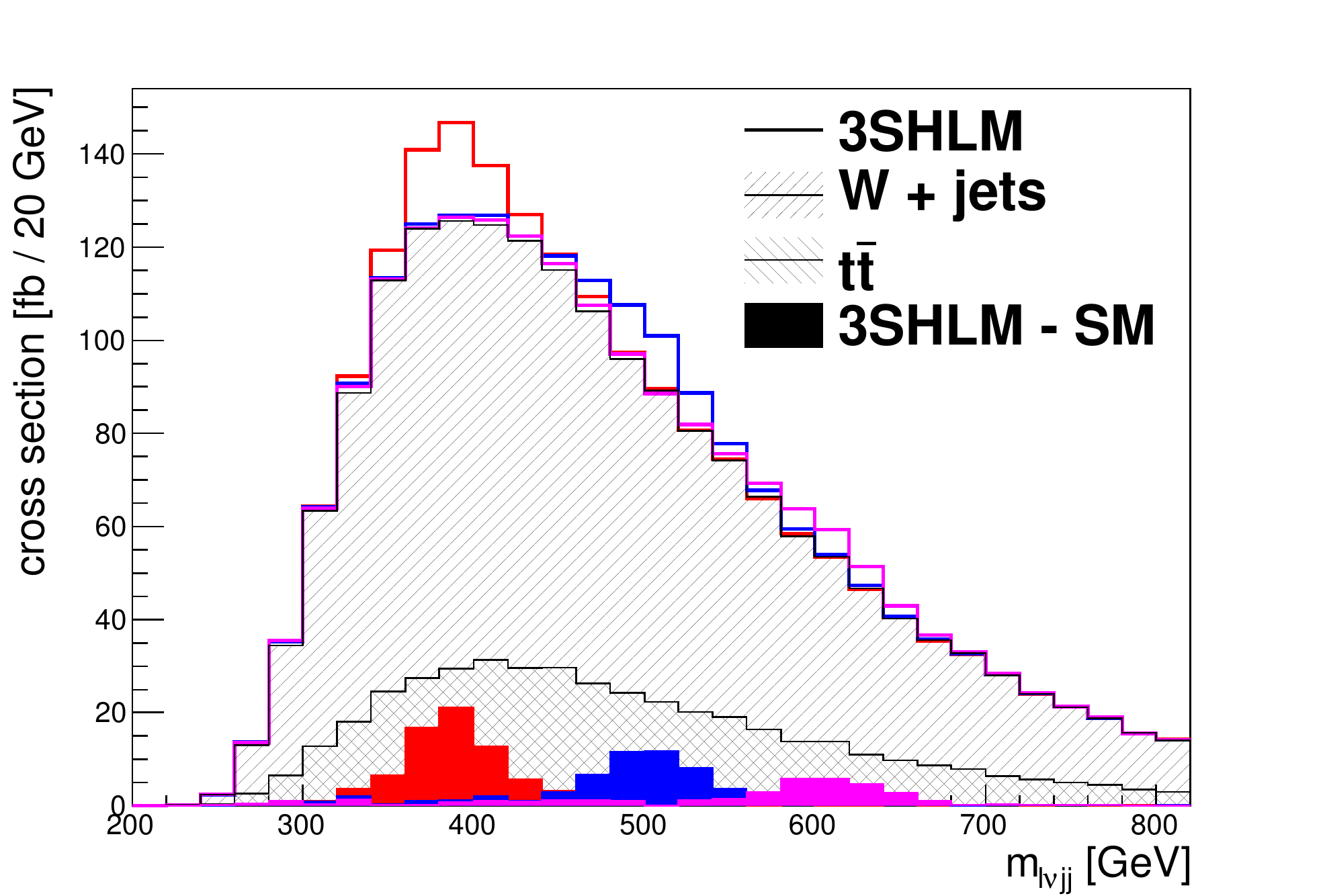}
 \includegraphics[scale=0.35]{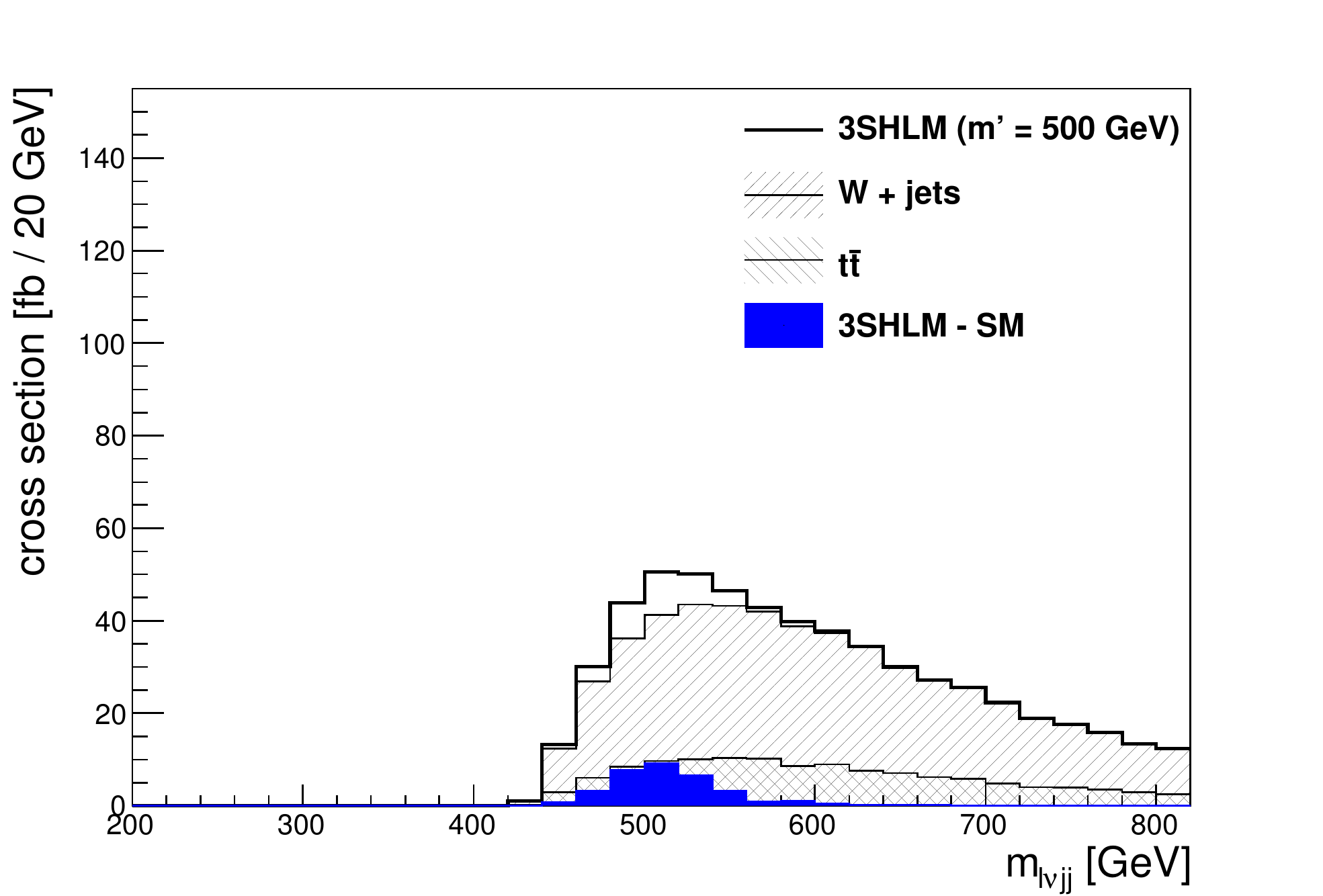}
 \caption{Total invariant mass distribution of the final state $\ell\nu jj$
after all cuts without (l.h.s) and including (r.h.s.) the $p_T$ cut on the
SM gauge boson momenta, with maximal $g_{W^\prime ff}$ allowed in the one-loop
scenario and \mbox{$m^\prime=380$} (red), 500 (blue) and \mbox{$600\,$GeV}
(pink).\label{lvjj_hWhZ}}
\end{figure*}

\subsubsection{Reconstruction of the Neutrino Momentum}

The momentum of the neutrino \mbox{$p_\nu=p$} is reconstructed from the
charged lepton 4-momentum \mbox{$p_\ell=q$} in the usual way by identifying
$\slashed{E}_T$ with the neutrino $p_T$ and using the on-shell condition for
the the intermediate $W$ boson
\begin{equation}\label{W_onshell}
 \left( p + q \right)^2 = m^2_W
\end{equation}
with massless leptons to account for the missing information. In general,
this quadratic equation has two solutions,
\begin{equation}
 p_z = \frac{ q_z \left( m_W^2 + 2 \vec{p}_T \cdot \vec{q}_T \right) \pm q_0 \sqrt{D} }{ 2 q_T^2 }
\end{equation}
with the determinant
\begin{equation}
 D = \left( m^2_W + 2 \vec{p}_T \cdot \vec{q}_T \right)^2 - 4 p^2_T q_T^2\;,\nonumber
\end{equation}
so that we pick the solution with a lower $\left| p_z \right|$, following
the argument \mbox{in~\cite{Abe:2011qe}}. However, due to off-shell effects
of the intermediate $W$ and detector smearing, a reasonable amount of events
(roughly $10\,\%$ at parton level and $25\,\%$ at detector level) fails to
give a real solution altogether, namely when $D$ becomes negative. In this
case, in order minimize the loss of statistics, we use \mbox{$D=0$} as
another condition and additionally solve for the invariant mass $m_{\ell\nu}$
of the leptonic $W$ in the resulting equations, finally cutting on
$m_{\ell\nu}$ corresponding to the mass cut in \mbox{eqn.~(\ref{Wcut_jj})}
to sort out unphysical results (\mbox{cf.~\cite{Bach:2009zz}} for a more
detailed  discussion of the method). Numerically it turns out that roughly
\mbox{$10\,\%$} of the original number of events can be reconstructed this
way with a competitive signal-to-background ratio in the resonance region,
so that they are included in the analysis. Ultimately, having obtained a
neutrino momentum and a corresponding $W$ boson momentum, we also apply
the $p_T$ and back-to-back \mbox{cuts~(\ref{WZ_btb})} on the leptonic and
the hadronic resonances before computing the total invariant mass
$m_{\ell\nu jj}$ of the final state (\mbox{cf.~fig.~\ref{lvjj_hWhZ}}).

\subsubsection{Disentangling the Jet Resonances}

As mentioned above, an important feature of the $\ell\nu jj$ final state
is that it encompasses the decay of both heavy gauge bosons with the same
signature, so that the only means of signal discrimination is to disentangle
the two SM gauge bosons in the di-jet resonance. At detector level, these
resonances have widths of the order of the mass splitting itself, which
makes it almost impossible to separate them merely by invariant mass cuts.
This problem was addressed at parton level \mbox{in~\cite{Ohl:2008ri}},
proposing a statistical method to numerically separate the two SM resonances
and hence the two heavy resonances: The true gauge boson counts $N_W$ and
$N_Z$ within a given sample are assumed to be smeared over a certain invariant
mass range according to underlying probability density functions $p_i(m)$,
which can be approximated by a convolution of the intrinsic Lorentz
distribution with an experimental detector response function described
by a Gaussian distribution. If the $p_i(m)$ are known at detector level,
the signal mixture
\begin{equation}
 \tilde{N}_i = \sum_j T_{ij} N_j\;, \qquad i,j = W,Z
\end{equation}
with
\begin{equation}\label{Tij}
 T_{ij} = \int_{L_i}^{U_i} \text{d}m\;p_j(m)
\end{equation}
can be inverted numerically to infer the true numbers $N_W$ and $N_Z$ from
the smeared numbers $\tilde{N}_W$ and $\tilde{N}_Z$ counted inside the
mass windows
\begin{align}\label{Li_Ui}
 L_W &= 60\,\text{GeV}\;, \nonumber \\
 U_W &= L_Z = \frac{m_W+m_Z}{2}\;, \nonumber \\
 U_Z &= 111\,\text{GeV}\;.
\end{align}

\begin{figure*}
 \includegraphics[scale=0.35]{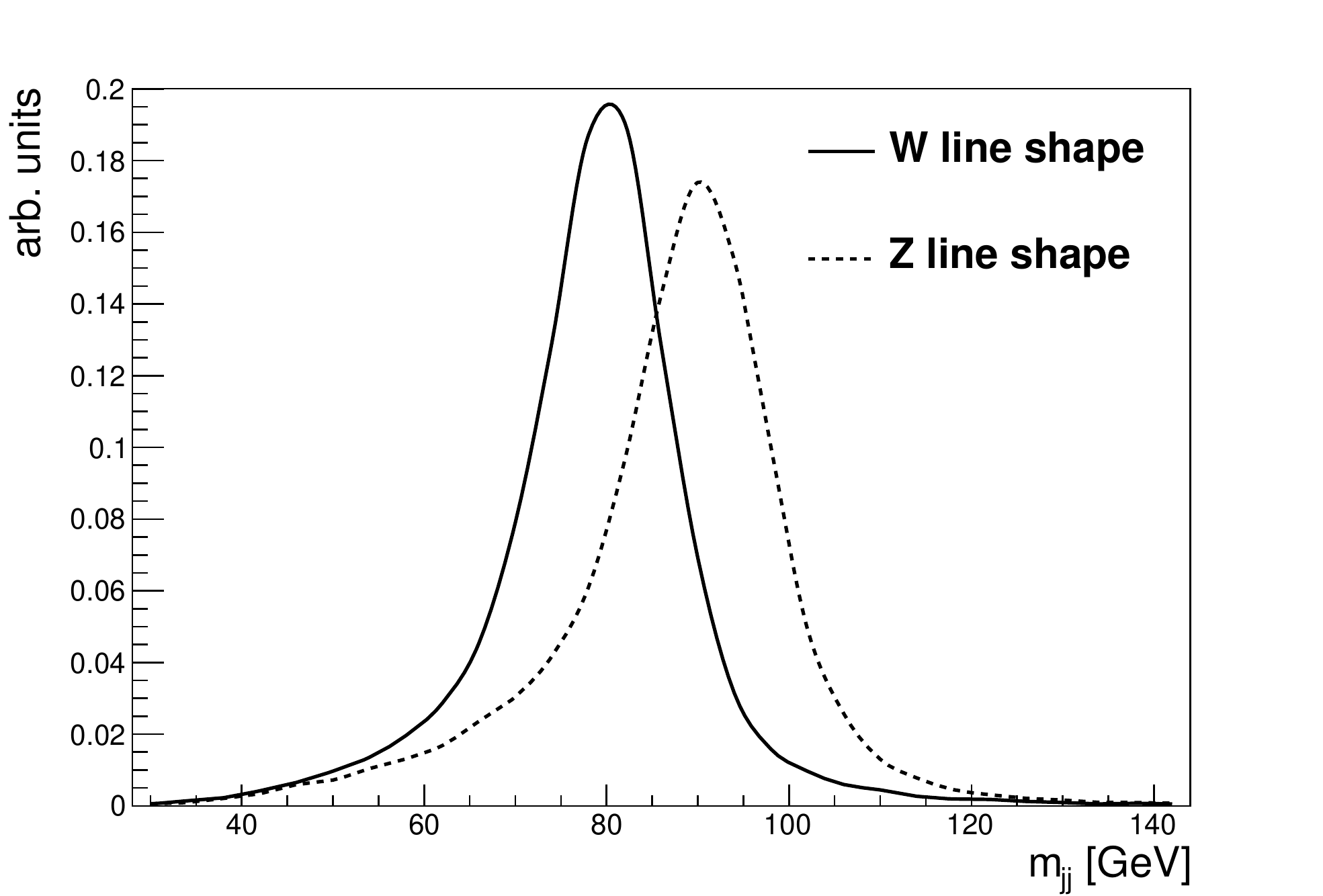}
 \includegraphics[scale=0.35]{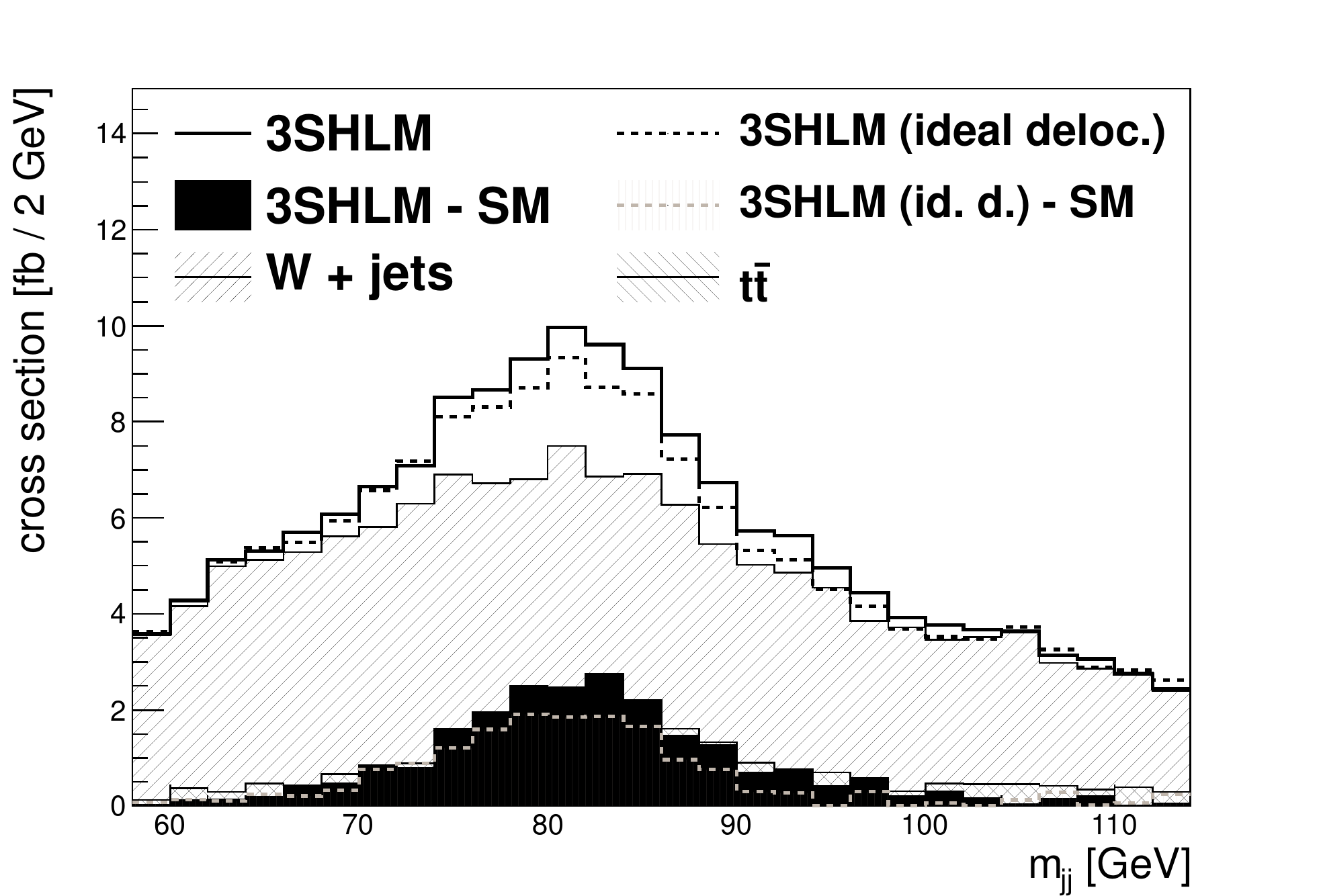}
 \caption{Normalized line shapes of the di-jet resonances of boosted $W$
and $Z$ bosons at detector level (l.h.s). The r.h.s. shows the invariant
di-jet mass distribution within the range of the SM bosons in the physical
samples, after cutting on the resonant region around $m^\prime$ in
$m_{\ell\nu jj}$.\label{WZ_fit}}
\end{figure*}

Whereas the authors of~\cite{Ohl:2008ri} assume an experimental Gaussian
smearing of \mbox{$10\,$GeV} as an input in their parton level study,
a different approach is chosen here: MC samples are produced at parton
level which contain strongly boosted back-to-back $W$ and $Z$ bosons
decaying hadronically, thus mimicking the final state topology of the
signal. Then the clean $W$ resp. $Z$ resonance line shapes in the invariant
di-jet mass obtained from these samples at detector level
(\mbox{cf.~fig.~\ref{WZ_fit}}) are used as an approximation for the unknown $p_i(m)$,
from which the mixing matrix $T$ and its inversion are then computed
numerically using \mbox{eqn.~(\ref{Tij})}, giving
\begin{equation}
 T^{-1} \approx \left( \begin{array}{cc} 1.68 & -0.82 \\ -0.60 & 1.89 \end{array} \right)\;.
\end{equation}

Having determined the $T^{-1}_{ij}$, the relevant signal events are isolated by
cutting on the total invariant mass  $m_{\ell\nu jj}$ within a region of
\mbox{$\pm30\,$GeV} around the heavy resonance
(\mbox{cf.~fig.~\ref{lvjj_hWhZ}}), where \mbox{$m^\prime=500\,$GeV} was
chosen here for a feasibility test of the method, because a medium value
of $m^\prime$ allows for the largest absolute values of $g_{W^\prime ff}$
(\mbox{cf.~\cite{Abe:2008hb}}) and hence possesses the highest discovery
potential. After this cut, the SM gauge boson resonances are examined in
the invariant mass distributions $m_{jj}$ of the corresponding jet pairs
(\mbox{cf.~fig.~\ref{WZ_fit}}) in order to obtain the mixed numbers
$\tilde{N}_i$ inside the cut windows $\{L_i,U_i\}$ specified in
\mbox{eqn.~(\ref{Li_Ui})}, and finally apply $T^{-1}$. The whole procedure
is carried out both with maximal $g_{W^\prime ff}$ as well as in ideal
delocalization for a cross check.

\begin{table}
\begin{tabular}{|c||c|c|c|c|}
 \hline
 \multicolumn{5}{|c|}{ideal delocalization}          \\
 \hline\hline
 $i$ & $\tilde{N}_i$ & $\tilde{s}_i$ & $N_i$ & $s_i$ \\
 \hline
 $W$ &          1203 &           6.8 &  1713 &  5.3  \\
 $Z$ &           373 &           2.5 &   -21 & -0.07 \\
 \hline
\end{tabular}
\hspace{0.03\textwidth}
\begin{tabular}{|c||c|c|c|c|}
 \hline
 \multicolumn{5}{|c|}{maximal $g_{W^\prime ff}$}     \\
 \hline\hline
 $i$ & $\tilde{N}_i$ & $\tilde{s}_i$ & $N_i$ & $s_i$ \\
 \hline
 $W$ &          1533 &           8.6 &  1972 & 6.1   \\
 $Z$ &           734 &           4.9 &   462 & 1.5   \\
 \hline
\end{tabular}
\caption{Signal events and significances as computed from
\mbox{eqn.~(\ref{def_sign})} for \mbox{$\int L=100\,$fb$^{-1}$} before
and after disentangling the signal as described in the text, in ideal
delocalization (left) and with maximal $W^\prime$ coupling to SM fermions
(right). The significance also drops for the $W$ signals because the
uncertainty generally becomes larger in the disentangling procedure.\label{WZ_sep}}
\end{table}

As detailed in \mbox{table~\ref{WZ_sep}}, the $Z$ significance, implying
the decay of a $W^\prime$, does indeed drop to nearly zero in ideal
delocalization as expected, while a finite significance \mbox{$s=1.5$}
remains with maximal $g_{W^\prime ff}$. In the parton level analysis
of~\cite{Ohl:2008ri} a significance \mbox{$s\sim2$} was found for the
disentangled $Z$ signal, whereas the basic differences between their
analysis and this one are that the detector width of SM bosons in jet pairs
was somewhat overestimated with 10 GeV, while, on the other hand, the
considerably large background contributions of inclusive jet production
at the detector were underestimated at parton level. In any case, this
result is still far away from a liable discovery threshold for an integrated
luminosity of \mbox{$100\,$fb$^{-1}$} considered here, so that the discovery
prospects for the $W^\prime$ remain very poor also in the $\ell\nu jj$
channel.

\subsection{The $\ell\nu\ell\ell$ Channel}\label{lvll}

\begin{figure*}
 \includegraphics[scale=0.35]{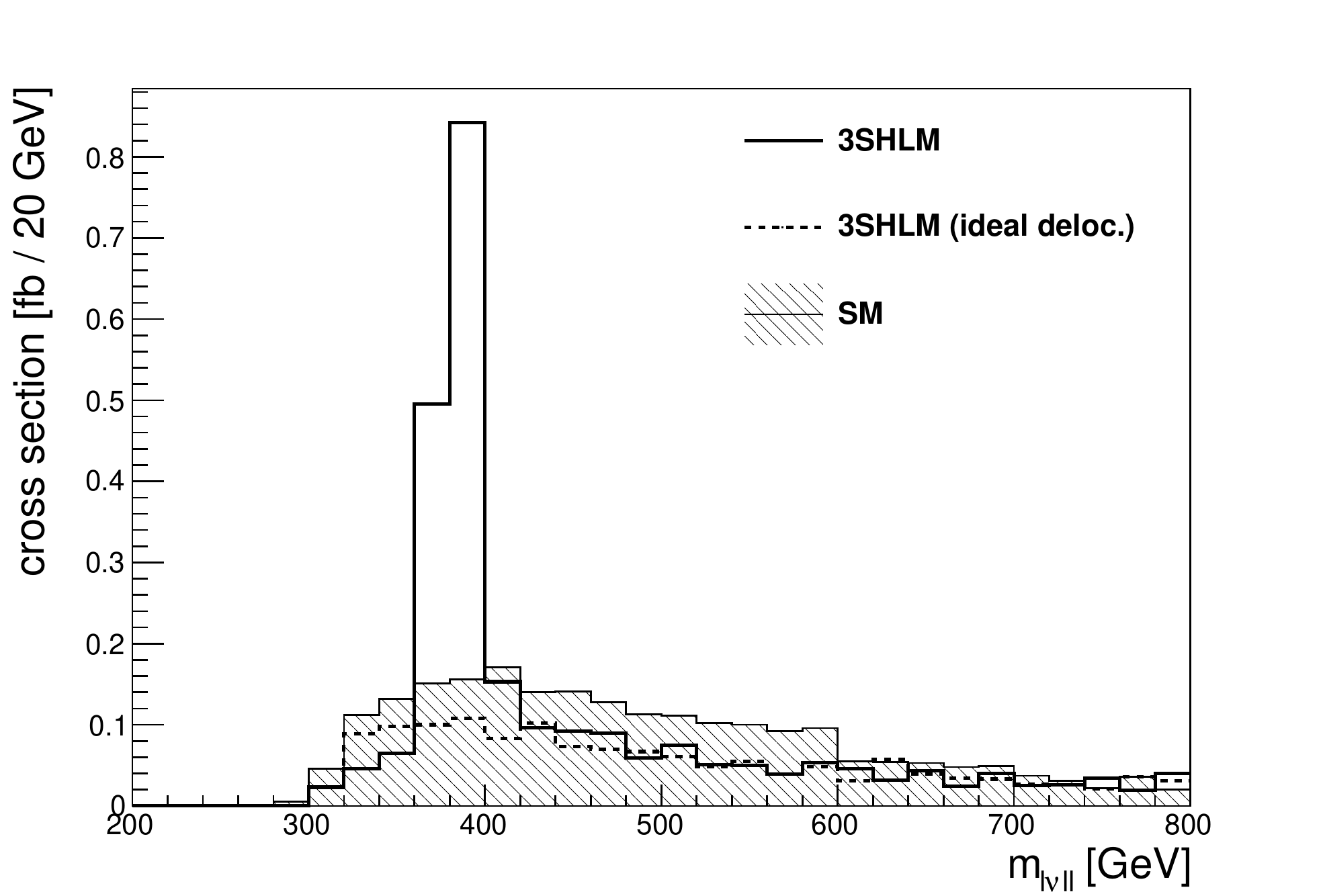}
 \includegraphics[scale=0.35]{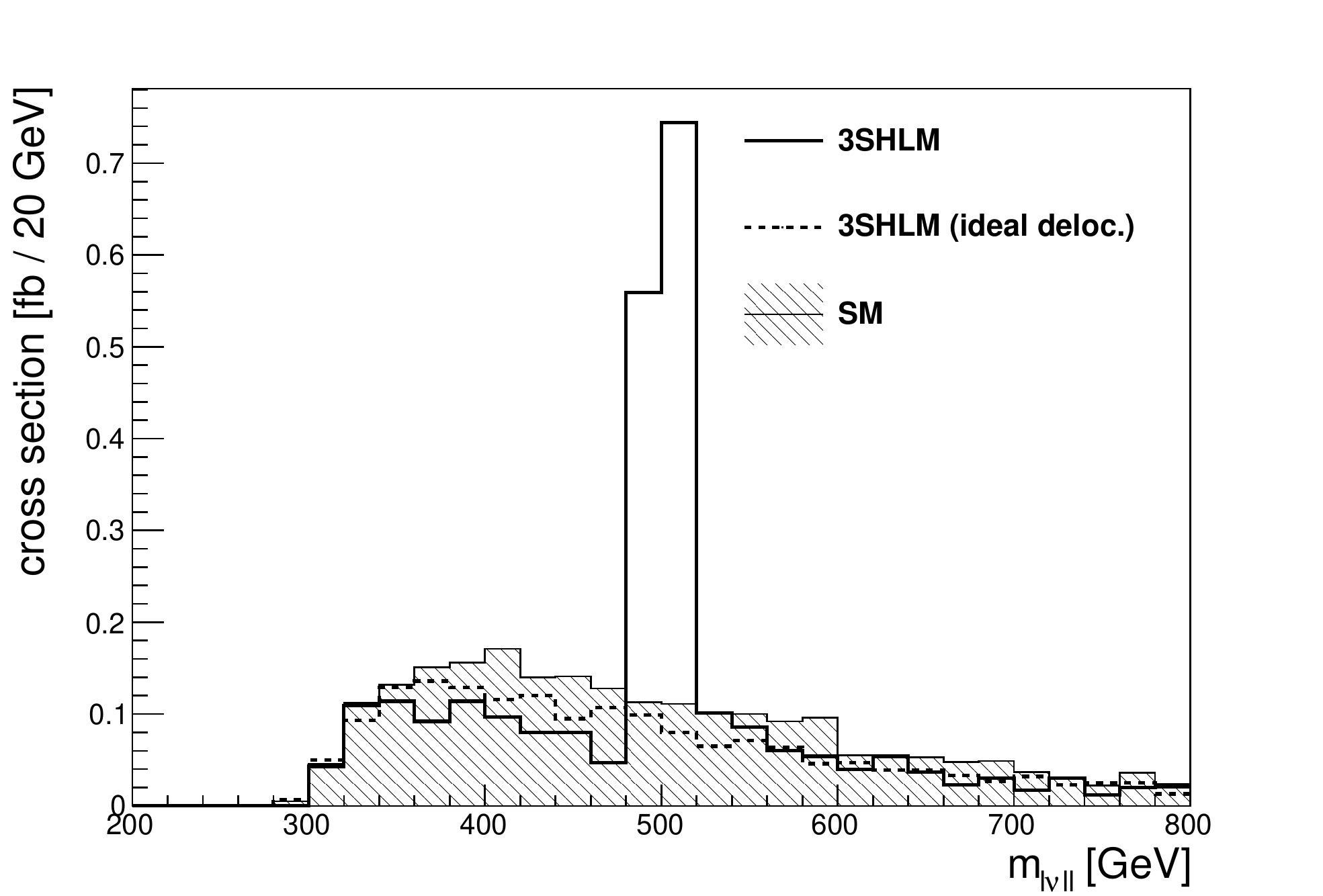}
 \includegraphics[scale=0.35]{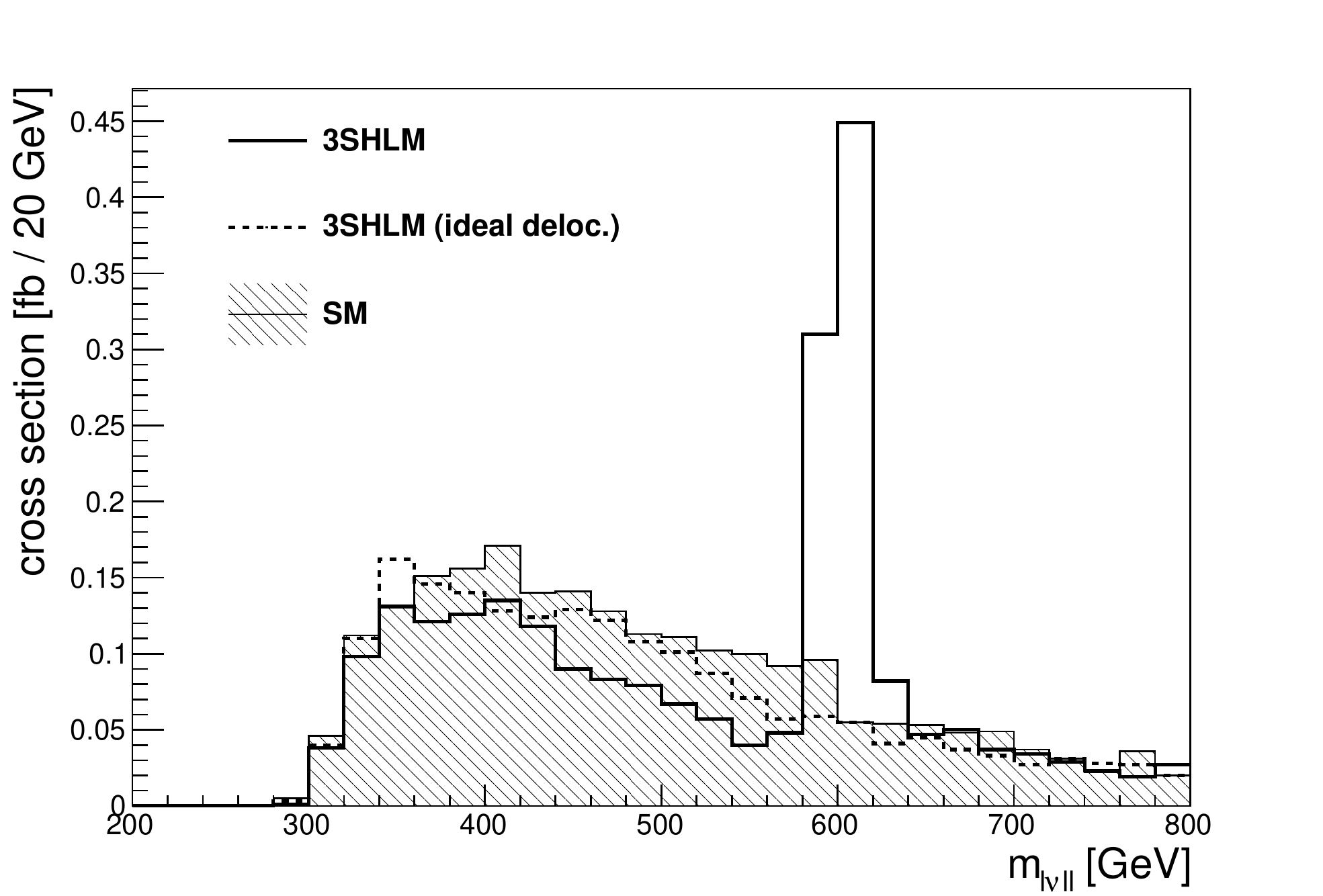}
 \caption{Total invariant mass of the final state $\ell\nu\ell\ell$ in the
SM as well as in the 3SHLM (ideal and non-ideal delocalization with max.
$g_{W^\prime ff}$ allowed) for \mbox{$m^\prime=380$}, 500 and
\mbox{600$\,$GeV} (parton level).\label{lnll_part}}
 \includegraphics[scale=0.35]{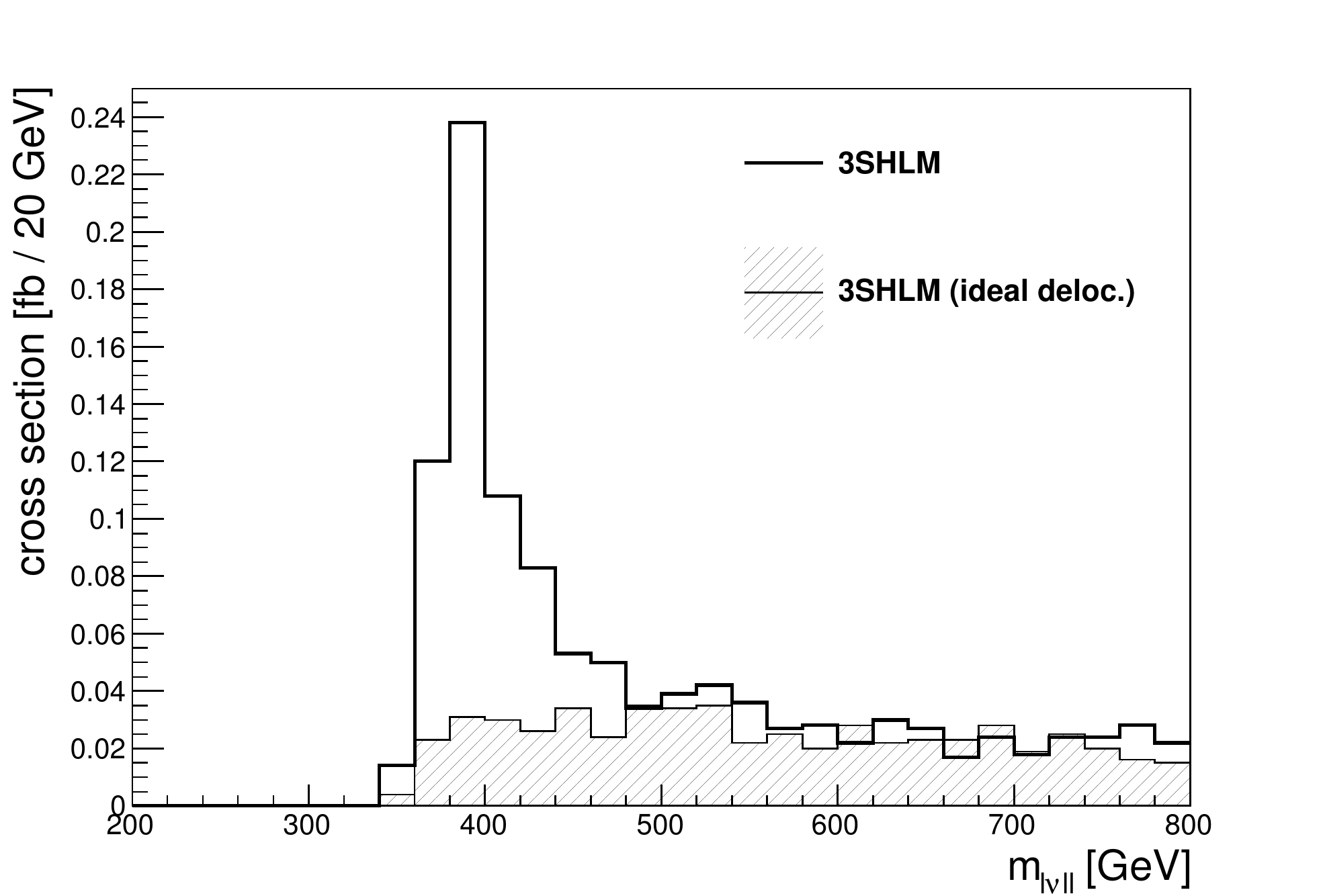}
 \includegraphics[scale=0.35]{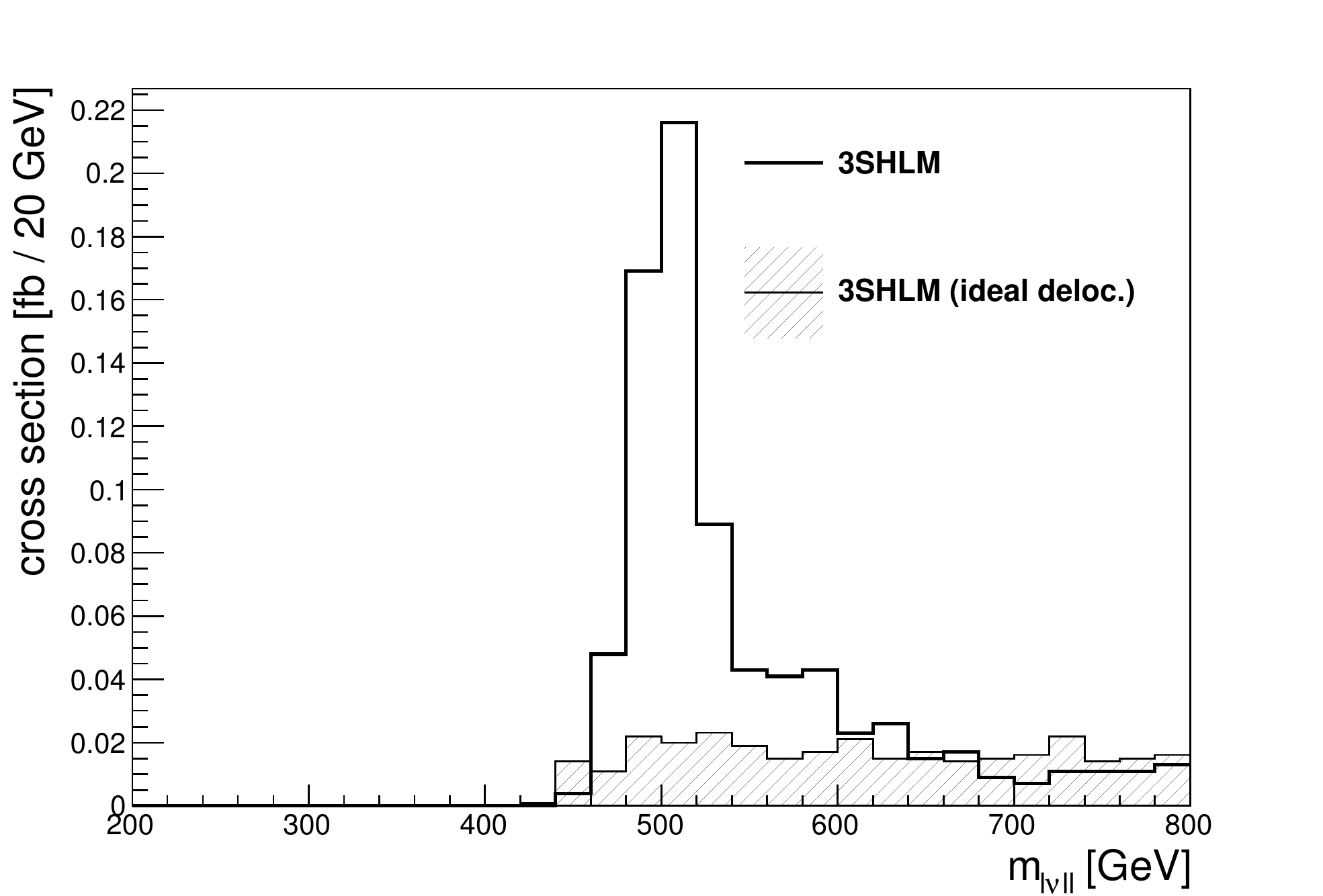}
 \includegraphics[scale=0.35]{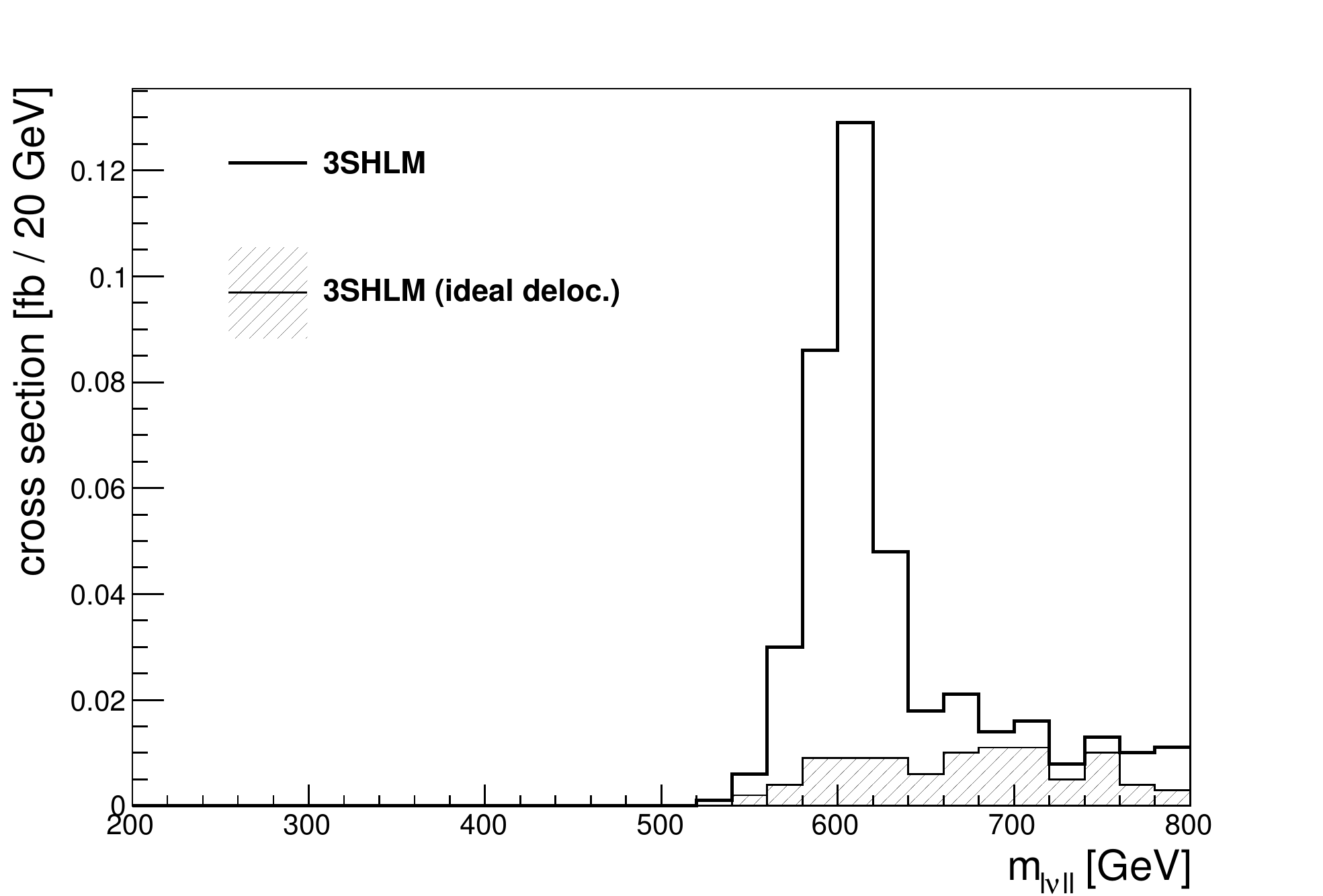}
 \caption{Total invariant mass of the final state $\ell\nu\ell\ell$ in the
3SHLM (ideal vs. non-ideal delocalization with max. $g_{W^\prime ff}$
allowed) for \mbox{$m^\prime=380$}, 500 and \mbox{600$\,$GeV}, including
the $p_T$ and back-to-back cuts on SM gauge \mbox{bosons~(\ref{WZ_btb})}
(detector level).\label{lnll_det}}
\end{figure*}

Another venue for a discriminative search for the heavy $W^\prime$ is the
purely leptonic final state $\ell\nu\ell\ell$, which possesses by far the
smallest cross section but at the same time a very clean detector signature
compared to the semileptonic channels considered above. The event selection
requirements are in this case exactly three isolated charged leptons passing
the kinematic cuts in \mbox{eqn.~(\ref{kincuts})} as well as missing
transverse momentum passing the $p_T$ cut. The further procedure simply
is to combine the treatments of one or two charged leptons discussed in
the previous sections: With mixed flavors, the lepton pair of equal flavor
is required to have opposite charges and the kinematic topology
corresponding to the decay of a boosted $Z$ boson, \mbox{i.\,e.} pass the
invariant mass cut pointed out in \mbox{eqn.~(\ref{Zcut_ll})}, while the
third lepton of different flavor is required to reconstruct at least one
neutrino momentum together with an $\slashed{E}_T$ (cf.~the discussion in
\mbox{section~\ref{lvjj}}) that also passes the kinematic cuts. With three
leptons of equal flavor, the procedure is to demand mixed charges and pick
from the two possible pairs of oppositely charged leptons the one whose
invariant mass is closer to the $Z$ mass, whereas the remaining lepton is
then required to produce at least one reasonable neutrino momentum together
with $\slashed{E}_T$. Finally, the reconstructed SM gauge boson momenta
should in any case also pass the $p_T$ and back-to-back cuts given in
\mbox{eqn.~(\ref{WZ_btb})}.

In contrast to the other channels, in this case not only the signal but
also the whole background is purely electroweak, so that sizeable interference
terms with diagrams containing heavy bosons or fermions can be expected and
are indeed present, qualitatively affecting the sideband shape around the
resonance compared to the SM sample. Moreover, even in the ideal
delocalization regime the electroweak couplings deviate to some extent from
the SM values, which also affects the size of the electroweak background
processes. The overall effects are twofold (\mbox{cf.~fig.~\ref{lnll_part}}):
on one hand the total height of the background shoulder is reduced, whereas
on the other hand there is a pronounced interference in
the resonance region with a sign change at $m^\prime$. In order to compute
significances, it is therefore sensible to compare the signal samples with
respective samples generated in the ideal delocalization setup rather than
a SM sample, thus accounting for any model effects except for resonant
$W^\prime$ diagrams. This approach is justified by the fact pointed out
above that, in any case, the $Z^\prime$ must be assumed to have
already been discovered at the eve of a 3SHLM-like $W^\prime$ search.

\begin{figure*}
 \includegraphics[scale=0.35]{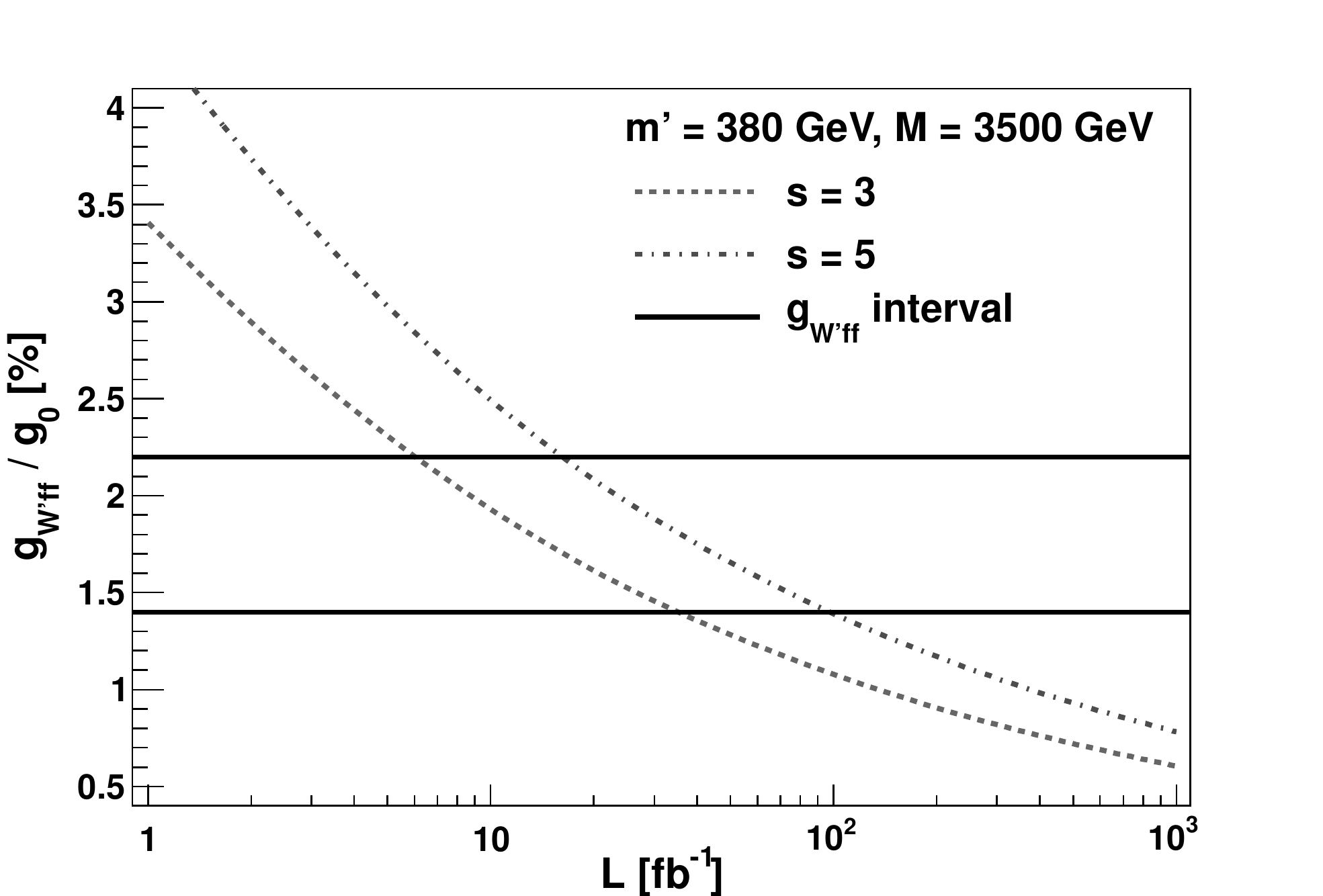}
 \includegraphics[scale=0.35]{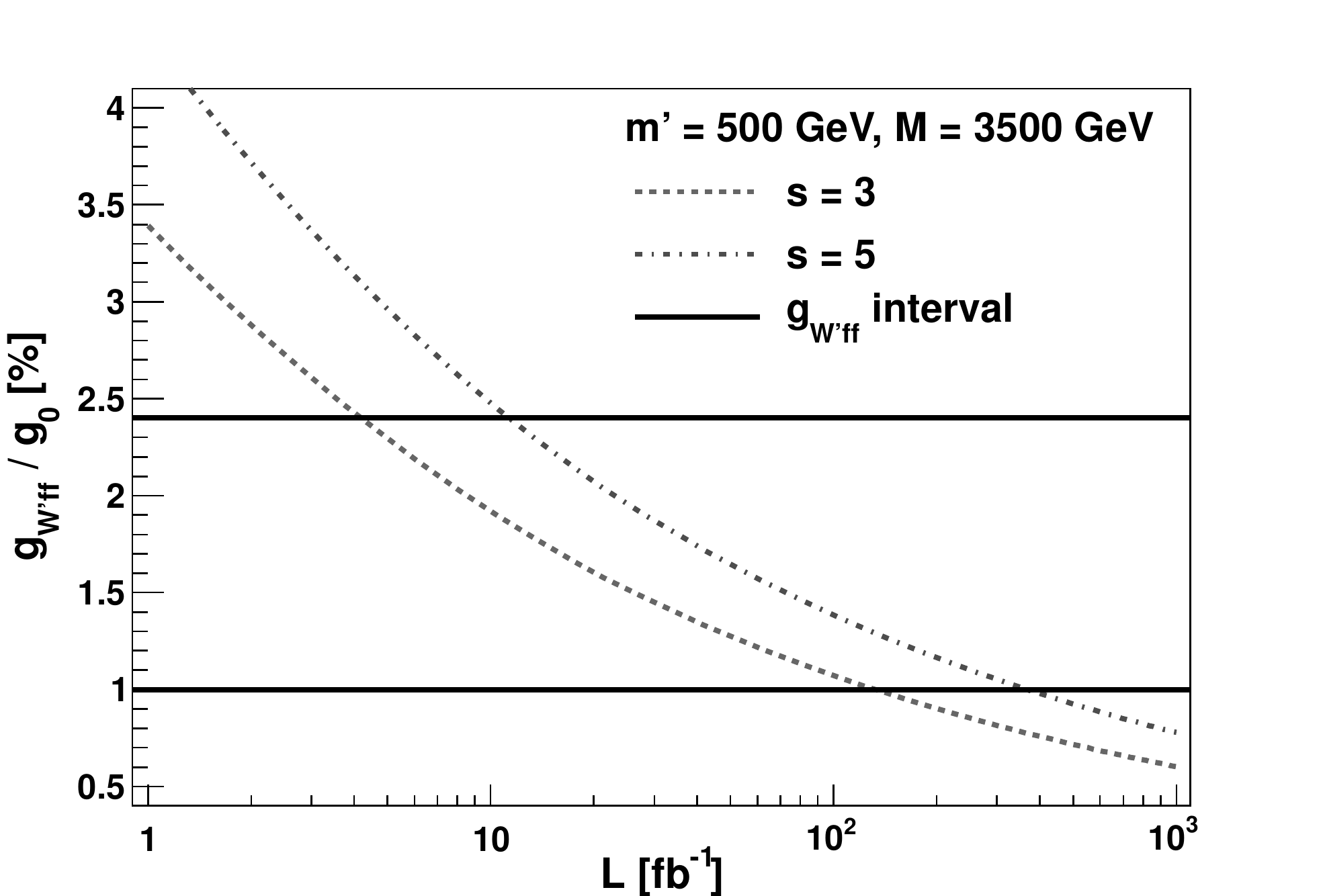}
 \includegraphics[scale=0.35]{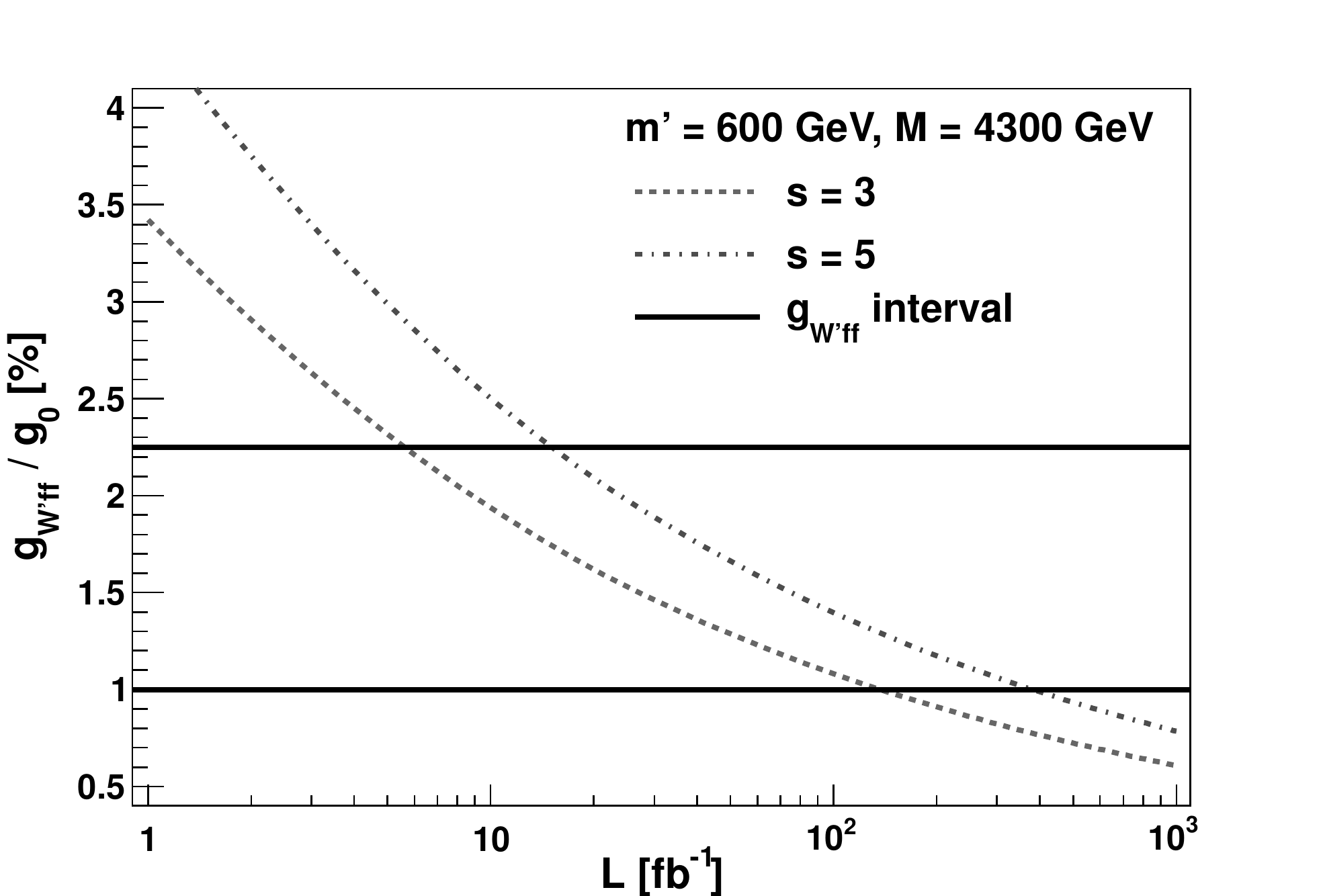}
 \caption{\mbox{$s=3$} and \mbox{$s=5$} sensitivity contours for a direct
$W^\prime$ search in the $\ell\nu\ell\ell$ channel at detector level,
depending on the available integrated luminosity and the $g_{W^\prime ff}$
value.
The values of $M$ have been chosen as low as possible while complying with
the bounds of~\cite{Abe:2008hb} and demanding that the heavy fermions remain
undetected in their respective search channels~\cite{Speckner:2010zi}.
Increasing $M$ generally shifts all $g_{W^\prime ff}$ windows to larger
values, thus improving detection as well as exlusion prospects.
\label{lnll_cont}}
\end{figure*}

As illustrated in \mbox{fig.~\ref{lnll_cont}}, despite its overall tiny cross
sections the purely leptonic channel turns out to be the most promising one
for a direct $W^\prime$ search compared to the results of the semileptonic
channels stated in the previous sections. In fact, depending on the point
chosen in parameter space, one of the multi-purpose experiments at the LHC
might be $5\sigma$-sensitive to the $W^\prime$ as soon as an integrated
luminosity of about \mbox{$\int L=10\,$fb$^{-1}$} with
\mbox{$\sqrt{s}=14\,$TeV} is collected (\mbox{$m^\prime=500\,$GeV}, max.
$g_{W^\prime ff}$ allowed). On the other hand, \mbox{fig.~\ref{lnll_cont}}
also implies that \mbox{$\int L\lesssim150\,$fb$^{-1}$} would be necessary
to cover the entire allowed parameter space of the non-ideally delocalized
3SHLM with a direct $W^\prime$ search at 3$\sigma$ \mbox{c.~l.}
(\mbox{$m^\prime=600\,$GeV}, min. $g_{W^\prime ff}$ allowed).
Note that for \mbox{$m^\prime=380\,$GeV} the minimal allowed $g_{W^\prime ff}$
depends on the choice of $M$, which could be as low as
\mbox{$\sim 2\,$TeV}~\cite{Abe:2008hb}.
However, anything below the value of \mbox{$3.5\,$TeV} would make the heavy
fermions visible in the respective search channels~\cite{Speckner:2010zi}.
Therefore we only consider higher values.

\section{Summary}\label{summ}

Besides the Higgs mechanism, various different approaches exist to describe
the spontaneous breaking of the electroweak symmetry, for example through
the introduction of one (or more) additional compact space-time dimension
rather than a new field. New gauge degrees of freedom emerge naturally from
the necessity to gauge the whole 5D bulk, which can then be used to
dynamically break the symmetry, \mbox{e.\,g.} by assigning non-trivial
vacuum configurations to these gauge components. The Three-Site Higgsless
Model~\cite{Chivukula:2006cg} represents a minimal approach in this respect,
since the extra dimension is discretized on only three sites, thus
eliminating all higher Kaluza-Klein excitations except for the two lightest
sets from the spectrum. In this setup, the heavy copies of the $W$ and the
$Z$ constitute the most accessible signatures of the new physics.

Since discovery thresholds for the $Z^\prime$ of the 3SHLM in the ideal
delocalization scenario at the LHC with \mbox{$\sqrt{s}=14\,$TeV} have been
published recently~\cite{Abe:2011qe} (\mbox{cf.~also~\cite{Bach:2009zz}}),
this study is focussed on the discovery potential of the model-specific
$W^\prime$ in the non-ideally delocalized
scenario~\cite{Abe:2008hb,Ohl:2008ri}, using it as a discriminative signature
against other models with heavy neutral resonances as well as against the
ideally delocalized scenario of the 3SHLM, where the $s$-channel signal of
the $W^\prime$ must also vanish. To that end, the signal as well as dominating
backgrounds with final state signatures $\ell\ell jj$, $\ell\nu jj$ and
$\ell\nu\ell\ell$ (dropping the hadronic final state $jjjj$ because of the
huge QCD backgrounds) were generated with the parton-level MC event generator
WHIZARD~\cite{Kilian:2007gr}, showered with PYTHIA~\cite{Sjostrand:2006za}
and finally processed with the generic detector simulation software
DELPHES~\cite{Ovyn:2009tx} mimicking the setup of the ATLAS experiment
at the LHC.

All three channels were then examined with respect to their discovery
prospects of a $W^\prime$, also testing a statistical method introduced
in~\cite{Ohl:2008ri} to disentangle the $W^\prime$ from the degenerate
$Z^\prime$ in the $\ell\nu jj$ channel at the detector level. It was found
that the $\ell\ell jj$ channel provides such a poor signal-to-background
ratio that a discovery within a reasonable amount of LHC data appears to
be out of reach even at the most promising point in the parameter space of
the non-ideally delocalized scenario. This is even more true for the
$\ell\nu jj$ channel after applying the said method to numerically disentangle
the $W^\prime$ and $Z^\prime$ signals, although the fact that afterwards the
$W^\prime$ significance drops indeed to nearly zero in ideal delocalization
whereas a residual significance remains in non-ideal delocalization can be
considered as a proof-of-method.

In the end, despite its tiny overall cross sections, the $\ell\nu\ell\ell$
channel remains as the only one with realistic discovery prospects for the
$W^\prime$ in the non-ideally delocalized 3SHLM with sensible assumptions
on LHC runtime and luminosity. In fact, at the most promising parameter
point, \mbox{i.\,e.} \mbox{$m^\prime=500\,$GeV} and maximal $g_{W^\prime ff}$
allowed within the model scenario, a $5\sigma$ discovery might already be
possible with \mbox{$\int L\sim10\,$fb$^{-1}$}, which is indeed only little
more than the \mbox{$8\,$fb$^{-1}$} reported \mbox{in~\cite{Abe:2011qe}}
to be necessary for the discovery of the ideally delocalized $Z^\prime$
in the $\ell\nu jj$ channel.
We obtain comparable if somewhat larger amounts
of reducible QCD background.
However, the non-ideal delocalization also lowers the discovery threshold
for the total degenerate $W^\prime$/$Z^\prime$ signal in the $\ell\nu jj$
channel, so that in this scenario---with the most optimistic choice of the
parameters $m^\prime$ and $g_{W^\prime ff}$---\mbox{a respective} discovery
may already be expected around \mbox{$\int L\sim5\,$fb$^{-1}$} at
\mbox{$\sqrt{s}=14\,$TeV} according to the present study.

Since the coupling $g_{W^\prime ff}$ is also bounded from below in the
non-ideally delocalized scenario, it is in principle possible to completely
exclude this scenario via a direct $W^\prime$ search. It turns out that
\mbox{$\int L\lesssim150\,$fb$^{-1}$} of LHC data would be required for a
complete $3\sigma$ exclusion, so that in summary, on the basis of the
present study, an $s$-channel search for the $W^\prime$ with final state
$\ell\nu\ell\ell$ may be considered a feasible approach to cover the entire
parameter space of the non-ideally delocalized 3SHLM within a realistic LHC
runtime. Clearly, it could pay off to also look for different, possibly more
sensitive approaches for an exclusion of the 3SHLM in the presence of a
$Z^\prime$-like resonance, for example a precise enough determination of
the suppressed $Z^\prime$ coupling to SM fermions $g_{Z^\prime ff}$, which
is essentially fixed by $m^\prime$ within the 3SHLM. The extraction of
liable exclusion bounds in this case could be a topic of further studies.

\bibliography{references}

\end{document}